\documentclass[]{revtex4}

\usepackage{amssymb}
\usepackage{natbib}
\usepackage{graphicx}
\usepackage{bm}
\usepackage{amsfonts}
\usepackage{amssymb}
\usepackage{amsmath}

\begin{document}
\title{Comments On "Three Paradox of Quantum Information"}
\author{wang hongyu\\Anshan Normal College,China}
\email{wanghy@mail.asnc.edu.cn}
\date{\today}
\begin{abstract}
The are some concept and calculate errors in "Three Paradoxes of
Quantum Information". When correct them,There will be no errors.
\end{abstract}
\maketitle

\section{comments}
When we say one state is a Entanglement-State,it means that the
state can not be factorized to a product of only two states of
subsytems.So when one want apply an operator on an entanglement
state,he cannot only factorized the state but should calculate it
directly.

As the paper,one can found that
$$H|0>=\frac 1 {\sqrt{2}} \{|0>+|1>\}$$
$$H|1>=\frac 1 {\sqrt{2}} \{|0>-|1>\}$$
Then
$$H \frac 1 {\sqrt{2}} \{|0>+|1>\}=|0>$$
But $\frac 1{\sqrt{2}}\{|00>+|11>\}$ don't equal to $\frac 1
{\sqrt{2}}\{|0>+|1>\}\bigotimes|\psi_2>$,so when one applied H on
the state $|00>+|11>$ for the first particle,it should be
\begin{eqnarray}
\nonumber H\frac 1 {\sqrt{s}}\{|00>+|11>\}=\frac 1 {\sqrt{2}}\{
(H|0>)|0>+(H|1>)|1>\}\\
\nonumber =\frac 1 2 \{|00>+|10>+|01>-|11>\}
\end{eqnarray}
Then applied H for the second particle,We got
$$H_1H_2 \frac 1 {\sqrt{2}} \{|00>+|11>\}=\frac 1 {\sqrt{2}}
\{|00>+|11>\}$$

and when follow this calculate,one would found there are no
paradoxes.

\end{document}